\documentclass[fleqn,twoside]{article}
\usepackage{espcrc2}

\usepackage{graphicx}

\title{Fast photon detection for the COMPASS RICH detector}

\author{
P.Abbon\address[CEA]
{\it CEA Saclay, DSM/DAPNIA, Gif-sur-Yvette, France\\[-0.3cm]}, 
M.Alekseev\address[Torino]
{\it INFN, Sezione di Torino and University of Torino, Torino, Italy\\[-0.3cm]}, 
H.Angerer\address[TUM]
{\it Technische Universit\"at M\"unchen, Physik Department, Garching, Germany\\[-0.3cm]}, 
M.Apollonio\address[Trieste]
{\it INFN, Sezione di Trieste and University of Trieste, Trieste, Italy\\[-0.3cm]}, 
R.Birsa\addressmark[Trieste], 
P.Bordalo\address[LIP]
{\it LIP, Lisbon, Portugal\\[-0.3cm]}, 
F.Bradamante\addressmark[Trieste], 
A.Bressan\addressmark[Trieste], 
L.Busso\addressmark[Torino], 
M.Chiosso\addressmark[Torino], 
P.Ciliberti\addressmark[Trieste],
M.L.Colantoni\address[Alles]
{\it INFN, Sezione di Torino and University of East Piemonte, Alessandria, Italy\\[-0.3cm]}, 
S.Costa\addressmark[Torino], 
S.Dalla Torre\addressmark[Trieste], 
T.Dafni\addressmark[CEA], 
E.Delagnes\addressmark[CEA], 
H.Deschamps\addressmark[CEA], 
V.Diaz\addressmark[Trieste], 
N.Dibiase\addressmark[Torino], 
V.Duic\addressmark[Trieste], 
W.Eyrich\address[Erlangen]
{\it Universit\"at Erlangen-N\"urnberg, Physikalisches Institut, Erlangen, Germany\\[-0.3cm]}, 
D.Faso\addressmark[Torino], 
A.Ferrero\addressmark[Torino], 
M.Finger\address[Prag]
{\it Charles University, Prague, Czech Republic and JINR, Dubna, Russia\\[-0.3cm]}, 
M.Finger Jr\addressmark[Prag], 
H.Fischer\address[Freiburg]
{\it Universit\"at Freiburg, Physikalisches Institut, Freiburg, Germany\\[-0.3cm]}, 
S.Gerassimov\addressmark[TUM], 
M.Giorgi\addressmark[Trieste], 
B.Gobbo\addressmark[Trieste], 
R.Hagemann\addressmark[Freiburg], 
D.von~Harrach\address[Mainz]
{\it Universit\"at Mainz, Institut f\"ur Kernphysik, Mainz, Germany \\[-0.3cm]}, 
F.H.Heinsius\addressmark[Freiburg], 
R.Joosten\address[Bonn]
{\it Universit\"at Bonn, Helmholtz-Institut f\"ur Strahlen- und Kernphysik, Bonn, Germany\\[-0.3cm]}, 
B.Ketzer\addressmark[TUM], 
K.K\"onigsmann\addressmark[Freiburg], 
V.N.Kolosov\address[CERN]
{\it CERN, European Organization for Nuclear Research, Geneva, Switzerland\\[-0.3cm]} 
\thanks{On leave from IHEP Protvino, Russia.},
I.Konorov\addressmark[TUM],
D.Kramer\address[Liberec]
{\it Technical University of Liberec, Liberec, Czech Republic}, 
F.Kunne\addressmark[CEA], 
A.Lehmann\addressmark[Erlangen], 
S.Levorato\addressmark[Trieste], 
A.Maggiora\addressmark[Torino], 
A.Magnon\addressmark[CEA],
A.Mann\addressmark[TUM], 
A.Martin\addressmark[Trieste], 
G.Menon\addressmark[Trieste], 
A.Mutter\addressmark[Freiburg], 
O.N\"ahle\addressmark[Bonn], 
F.Nerling\addressmark[Freiburg],
D.Neyret\addressmark[CEA], 
P.Pagano\addressmark[Trieste], 
S.Panebianco\addressmark[CEA], 
D.Panzieri\addressmark[Alles], 
S.Paul\addressmark[TUM], 
G.Pesaro\addressmark[Trieste], 
J. Polak\addressmark[Liberec], 
P.Rebourgeard\addressmark[CEA], 
F.Robinet\addressmark[CEA], 
E.Rocco\addressmark[Trieste],
P.Schiavon\addressmark[Trieste], 
C.Schill\addressmark[Freiburg]
\thanks {Corresponding author, mail: \texttt{Christian.Schill@cern.ch}}, 
W.Schr\"oder\addressmark[Erlangen], 
L.Silva\addressmark[LIP], 
M.Slunecka\addressmark[Prag], 
F.Sozzi\addressmark[Trieste],
L.Steiger\addressmark[Prag], 
M.Sulc\addressmark[Liberec], 
M.Svec\addressmark[Liberec], 
F.Tessarotto\addressmark[Trieste], 
A.Teufel\addressmark[Erlangen], 
H.Wollny\addressmark[Freiburg] }

\begin{document}

\begin{abstract}  

Particle identification at high rates  is a central aspect of many present and
future experiments in high-energy particle physics. The COMPASS experiment at
the SPS accelerator at CERN uses a large scale Ring Imaging CHerenkov detector
(RICH) to identify pions, kaons and protons in a wide momentum range. For the
data taking in 2006, the COMPASS RICH has been upgraded in the central photon
detection area ($25$\% of the surface) with a new technology to detect
Cherenkov photons at very high count rates of several $10^6s^{-1}$ per channel
and a new dead-time free read-out system, which allows trigger rates up to
$100$~kHz. The Cherenkov photons are detected by an array of $576$ visible and
ultra-violet sensitive multi-anode photomultipliers with $16$ channels each.
Lens telescopes of    fused silica lenses have been designed and built to focus
the Cherenkov photons onto the individual photomultipliers. The read-out
electronics of the PMTs is based on the MAD4 amplifier-discriminator chip and
the dead-time free high resolution F1-TDC. The $120$~ps time resolution of the
digital card guarantees negligible background from uncorrelated physical
events.  In the outer part of the detector, where the particle rates are lower,
the present multi-wire proportional chambers (MWPC) with Cesium Iodide
photo-cathodes have been upgraded with a new read-out electronic system based
on the APV preamplifier and shaper ASIC with analog pipeline and sampling ADCs.
The project was fully designed and implemented in the period November 2004
until May 2006. The upgraded detector showed an excellent performance during
the 2006 data taking: the number of detected Cherenkov photons per ring was
increased from $14$ to above $60$ at saturation. The time resolution was
improved from about 3 microseconds to about one nanosecond which allows an
excellent suppression of the background  photons from uncorrelated events.

\end{abstract}

\maketitle

\section{Introduction}

The COMPASS experiment \cite{compass} at the CERN SPS investigates key issues
in hadron physics by scattering high-energy polarized muons and hadrons off
polarized and unpolarized solid state targets. The produced particles are
detected in a large size two-stage forward spectrometer which covers a wide
kinematic range ($10^{-5} < x_{Bj} <  0.5$, $10^{-3}<Q^2<100$~(GeV/c)$^2$). One of
the key features of the COMPASS spectrometer is an excellent particle
identification which allows to separate scattered electrons, muons, pions,
kaons and protons. It is performed by electromagnetic and hadron calorimeters,
muon walls and a large scale Ring Imaging Cherenkov detector (RICH)
\cite{albrecht}. The RICH detector uses $C_4F_{10}$ as radiator gas inside a $5
\times 6$~m$^2$ wide and $3$~meter deep vessel. The produced Cherenkov photons
are reflected by a $20$~m$^2$ mirror wall onto a set of photon-detectors: up to
2004 multi-wire proportional chambers (MWPC) with Cesium Iodide photo-cathodes,
covering a total active surface of about $5.5$~m$^2$ and including in total about
80.000 read-out channels.

\section{Motivation of the upgrade}  The read-out system of the COMPASS  RICH
photon detectors in use till 2004 consisted of Gassiplex front-end chips
\cite{gassiplex} connected to the MWPCs, which have an integration time of
about $3$~microseconds. In the experimental environment of the COMPASS setup
there is a large flux of Cherenkov photons especially in the centre of the
detector, since the CERN muon beam is accompanied by 10\% to 20\% of halo muons
which pass through the detector and create Cherenkov rings. At high beam
intensities of $0.4$ to $1.0\times 10^8$ muons per second, these halo rings 
create a considerable background of overlapping rings in the centre of the
detector, which reduces the particle identification efficiency and purity,
especially for particles in the very forward direction.

\begin{figure}[t]
\includegraphics[width=\columnwidth]{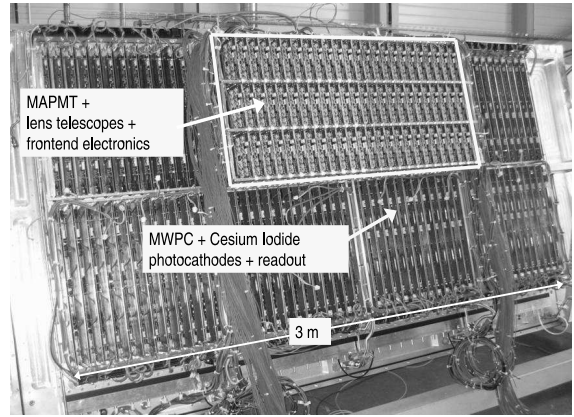}\\[-1.4cm]
\caption{Lower half of the COMPASS RICH photon detector before installation. 
The central part (marked by the white line) of the MWPCs has been replaced by 
lens telescopes, MAPMTs and fast read-out electronics consisting of MAD4
discriminators and F1 TDC.}
\label{detector}
\vspace{-0.5cm}
\end{figure}

A fast detection system for Cherenkov photons is needed to distinguish  by time
information the photons originating from scattered particles in the physics
events from the background from uncorrelated halo muons. The upgrade of the
COMPASS RICH detector consists of two parts. In the central part of the
detector (25\% of the sensitive area) where the photon flux is largest the
MWPCs with Cesium Iodide photo-cathodes have been replaced by multi-anode
photomultipliers \cite{mapmt} and a fast read-out system based on the MAD4
\cite{mad4} discriminator front-end chip and the dead-time free F1 TDC
\cite{tdc} (see Fig. \ref{detector}). In the outer part, the existing MWPC
chambers have been equipped with a faster read-out based on the APV preamplifier
with sampling ADCs \cite{apv}. In addition, the upgraded detector allows a high
rate operation at increased trigger rates from $20$ up to $100$~kHz and small
dead-time \cite{daq}.

\section{The upgrade of the central region} The central region upgrade is
performed by replacing the central part of the MWPC photon detector set by an
array of 576 UV-sensitive 16 channel multi-anode PMTs H7600-03-M16 from
Hamamatsu \cite{mapmt}. The sensitive wavelength range of these
photomultipliers extends from $200$~nm up to $700$~nm compared to the effective
sensitive range of the Cesium-Iodide photo-cathodes of $160$~nm to $200$~nm.
Therefore the number of detected photons per ring is expected to be about $4$
times larger. In addition, the ring resolution is improved by the larger number
of detected photons from $0.6$ to $0.3$~mrad. This will increase the upper
limit of the kinematic region of the $2 \sigma $ pion from  kaon separation
from $44$ to above $50$~GeV/c hadron momentum.

\begin{figure}[t]
\includegraphics[height=\columnwidth,angle=-90]{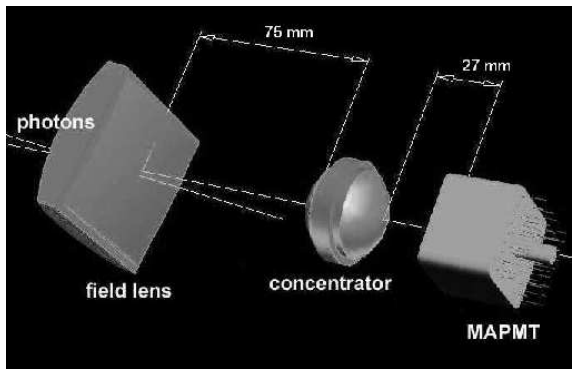}\\[-1.0cm]
\caption{Optical arrangement of the MAPMT and the fused silica lens telescope.}
\vspace{-0.5cm}
\label{lenses}
\end{figure}

The Cherenkov photons are focused onto the PMTs by 576 individual lens
telescopes from UV-transparent fused silica lenses (see Fig. \ref{lenses}). The
telescopes have been customly designed from one spherical and one aspherical lens
and offer a large angular acceptance for the photons of $\pm 9.5^o$ and  a
minimum image distortion. The image reduction is of a factor of about $7$ in
area; the PMT pixels are of $4.5 \times 4.5$~mm$^2$, the effective pixel size
of the new detector is $12\times 12$~mm$^2$. All PMTs have been shielded by
soft iron boxes to protect them from the residual $200$~Gauss field of the
$1$~Tesla open spectrometer magnet few meters away.

The PMTs are read out by frontend electronics mounted directly on the detector,
which consists of the MAD4 preamplifier and discriminator and the F1 TDC chip.
The MAD4 chip \cite{mad4} has a low noise level of $5-7$~fC compared to the
average PMT signal of about $500$~fC. It is able to handle signals up to
$1$~MHz rate. In $2007$, the MAD4 will be replaced by C-MAD chips, which have a
rate capability up to $5$~MHz. The digital part of the read-out consists of the
dead-time free F1 TDC \cite{tdc} mounted on the DREISAM\footnote{Digital
REad-out
Integrating and SAMpling card.} frontend card. It has a time resolution
of better than $120$~ps and can handle input rates up to $10$~MHz at trigger
rates of up to $100$~kHz. The data from one frontend card with $64$ read-out
channels are transferred via optical links to the COMPASS read-out system
\cite{daq}. All read-out electronics is mounted in a very compact setup
directly on the detector.

\begin{figure}[t] \includegraphics[width=\columnwidth]{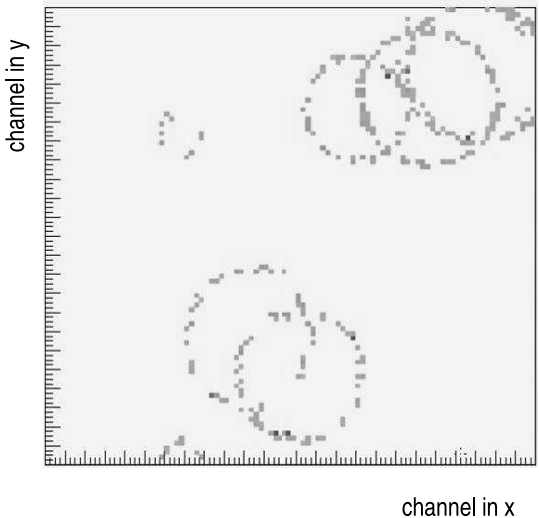}\\[-1.4cm]
\caption{Single physics event in the RICH detector, central region, after
applying a time cut of $5$~ns around the physics trigger. One can 
see Cherenkov rings of several hadrons.} \label{rings} \vspace{-0.8cm} \end{figure}

\begin{figure}[t]
\includegraphics[width=\columnwidth]{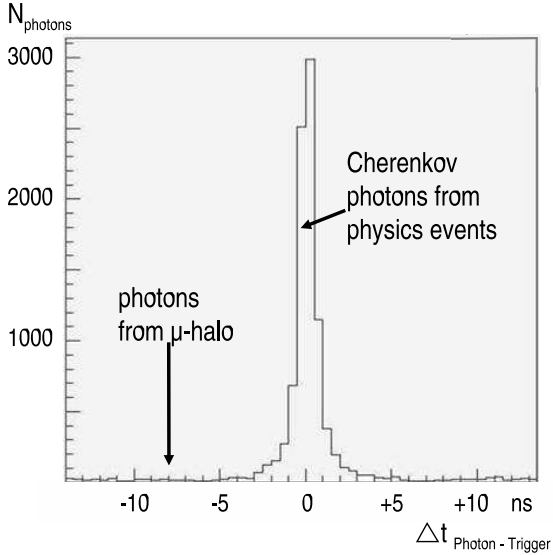}\\[-1.4cm]
\caption{Time distribution of the Cherenkov photons relative to the physics
trigger. One can see an excellent time resolution of about $1$~ns.}
\label{timing}
\vspace{-0.5cm}
\end{figure}

\begin{figure}[t]
\includegraphics[width=\columnwidth]{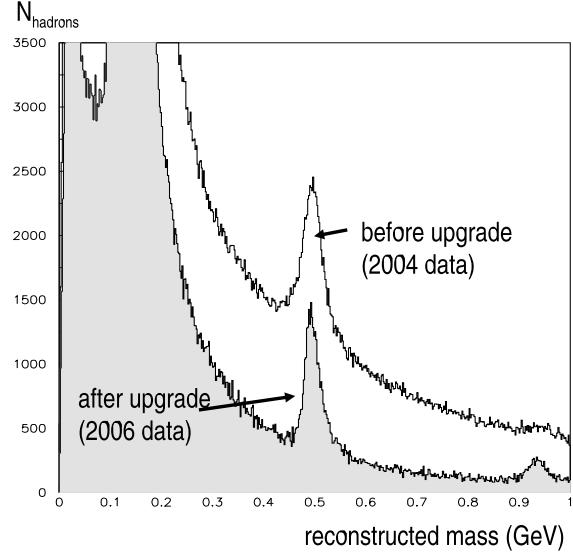}\\[-1.4cm]
\caption{RICH-reconstructed kaon mass-peak before (2004 data) and 
after (preliminary 2006 data) the RICH upgrade.}
\label{masspeak}
\vspace{-0.5cm}
\end{figure}

\section{Detector performance after the upgrade} In the COMPASS beam-time 2006,
all data have been taken with the upgraded RICH detector (see
Figs. \ref{rings}, \ref{timing} and \ref{masspeak}). The detector shows excellent performance: the number of
photons per ring has increased from $14$ before the upgrade to above $60$ at
saturation ($\beta \approx 1$). The increased statistics in the number of
photons has improved the ring resolution to $0.3$~mrad from $0.6$~mrad before.
While the increased number of photons has improved the particle identification
capability at lower momenta, the better ring resolution extends the particle
identification towards higher momenta. The time resolution for single photons
(see Fig. \ref{timing}) is about $1$~ns, which allows an almost complete
rejection of the background from uncorrelated halo muon Cherenkov signals.

\section{Conclusions}

A fast photon detection system with multi-anode photomultipliers and fast
read-out electronics based on the MAD4 discriminator and the F1 TDC has been
designed and constructed to upgrade the Ring Imaging Cerenkov (RICH) detector
of the COMPASS experiment. The upgraded detector was ready for the beam-time in
2006 and first data show an excellent performance of the new photon detection
system. The time resolution of about $1$~ns allows an almost complete rejection
of the background Cherenkov photons from uncorrelated muon halo events, which
improves efficiency and purity (see Fig. \ref{masspeak}) of the particle
identification especially in the very forward direction. The increased number
of detected photons extends the particle identification performance of the
COMPASS RICH both towards lower Cherenkov angles and at high particle momenta
above $50$~GeV/c.

\section*{Acknowledgments}
We acknowledge support from the CERN/PH groups PH/TA1,
TA2, DT2, TS/SU, and support by the BMBF (Germany) and the European Community-research
Infrastructure Activity under the FP6 program (Hadron Physics, RII3-CT-2004-506078).

\end{document}